%% file: paper.tex
\documentclass[11pt]{article}
\usepackage{epsfig}
\usepackage{amsmath}
\usepackage{hhline}
\usepackage{amssymb}

\newlength{\dinwidth}
\newlength{\dinmargin}
\begin{document}  
\newcommand{\pom}{{I\!\!P}}
\newcommand{\slowpi}{\pi_{\mathit{slow}}}
\newcommand{\fiidiii}{F_2^{D(3)}}
\newcommand{\fiidiiiarg}{\fiidiii\,(\beta,\,Q^2,\,x)}
\newcommand{\n}{1.19\pm 0.06 (stat.) \pm0.07 (syst.)}
\newcommand{\nz}{1.30\pm 0.08 (stat.)^{+0.08}_{-0.14} (syst.)}
\newcommand{\fiidiiiful}{F_2^{D(4)}\,(\beta,\,Q^2,\,x,\,t)}
\newcommand{\fiipom}{\tilde F_2^D}
\newcommand{\ALPHA}{1.10\pm0.03 (stat.) \pm0.04 (syst.)}
\newcommand{\ALPHAZ}{1.15\pm0.04 (stat.)^{+0.04}_{-0.07} (syst.)}
\newcommand{\fiipomarg}{\fiipom\,(\beta,\,Q^2)}
\newcommand{\pomflux}{f_{\pom / p}}
\newcommand{\nxpom}{1.19\pm 0.06 (stat.) \pm0.07 (syst.)}
\newcommand {\gapprox}
   {\raisebox{-0.7ex}{$\stackrel {\textstyle>}{\sim}$}}
\newcommand {\lapprox}
   {\raisebox{-0.7ex}{$\stackrel {\textstyle<}{\sim}$}}
\def\gsim{\,\lower.25ex\hbox{$\scriptstyle\sim$}\kern-1.30ex%
\raise 0.55ex\hbox{$\scriptstyle >$}\,}
\def\lsim{\,\lower.25ex\hbox{$\scriptstyle\sim$}\kern-1.30ex%
\raise 0.55ex\hbox{$\scriptstyle <$}\,}
\newcommand{\pomfluxarg}{f_{\pom / p}\,(x_\pom)}
\newcommand{\dsf}{\mbox{$F_2^{D(3)}$}}
\newcommand{\dsfva}{\mbox{$F_2^{D(3)}(\beta,Q^2,x_{I\!\!P})$}}
\newcommand{\dsfvb}{\mbox{$F_2^{D(3)}(\beta,Q^2,x)$}}
\newcommand{\dsfpom}{$F_2^{I\!\!P}$}
\newcommand{\gap}{\stackrel{>}{\sim}}
\newcommand{\lap}{\stackrel{<}{\sim}}
\newcommand{\fem}{$F_2^{em}$}
\newcommand{\tsnmp}{$\tilde{\sigma}_{NC}(e^{\mp})$}
\newcommand{\tsnm}{$\tilde{\sigma}_{NC}(e^-)$}
\newcommand{\tsnp}{$\tilde{\sigma}_{NC}(e^+)$}
\newcommand{\st}{$\star$}
\newcommand{\sst}{$\star \star$}
\newcommand{\ssst}{$\star \star \star$}
\newcommand{\sssst}{$\star \star \star \star$}
\newcommand{\tw}{\theta_W}
\newcommand{\sw}{\sin{\theta_W}}
\newcommand{\cw}{\cos{\theta_W}}
\newcommand{\sww}{\sin^2{\theta_W}}
\newcommand{\cww}{\cos^2{\theta_W}}
\newcommand{\trm}{m_{\perp}}
\newcommand{\trp}{p_{\perp}}
\newcommand{\trmm}{m_{\perp}^2}
\newcommand{\trpp}{p_{\perp}^2}
\newcommand{\alp}{\alpha_s}

\newcommand{\alps}{\alpha_s}
\newcommand{\sqrts}{$\sqrt{s}$}
\newcommand{\LO}{$O(\alpha_s^0)$}
\newcommand{\Oa}{$O(\alpha_s)$}
\newcommand{\Oaa}{$O(\alpha_s^2)$}
\newcommand{\PT}{p_{\perp}}
\newcommand{\JPSI}{J/\psi}
\newcommand{\sh}{\hat{s}}
\newcommand{\uh}{\hat{u}}
\newcommand{\MP}{m_{J/\psi}}
\newcommand{\PO}{I\!\!P}
\newcommand{\xbj}{x}
\newcommand{\xpom}{x_{\PO}}
\newcommand{\ttbs}{\char'134}
\newcommand{\xpomlo}{3\times10^{-4}}  
\newcommand{\xpomup}{0.05}  
%
%
\newcommand{\qsq}{\ensuremath{Q^2} }
\newcommand{\gevsq}{\ensuremath{\mathrm{GeV}^2} }
\newcommand{\et}{\ensuremath{E_t^*} }
\newcommand{\rap}{\ensuremath{\eta^*} }
\newcommand{\gp}{\ensuremath{\gamma^*p} }
\newcommand{\dsiget}{\ensuremath{{\rm d}\sigma_{ep}/{\rm d}E_t^*} }
\newcommand{\dsigrap}{\ensuremath{{\rm d}\sigma_{ep}/{\rm d}\eta^*} }

\begin{titlepage}
\begin{flushleft}
{\tt DESY 97-179}\hfill {\tt ISSN 0418-9833} \\
\end{flushleft}
\vspace*{3.0cm}
\begin{center}\begin{LARGE}
{\bf Low} ${\mathbf Q^2}$ {\bf Jet Production at HERA} \\
{\bf and Virtual Photon Structure} \\
\vspace*{2.5cm}
H1 Collaboration \\
\vspace*{2.5cm}
\end{LARGE}
{\bf Abstract}
\begin{quotation}
\noindent The transition between photoproduction and deep-inelastic
scattering is investigated in jet production at the HERA $ep$ collider, 
using data collected by the H1 experiment. Measurements of the 
differential inclusive jet cross-sections \dsiget and
\ensuremath{{\rm d}\sigma_{ep}/{\rm d}\eta^*},
where \et and \rap are the transverse energy and the pseudorapidity
of the jets in the virtual photon-proton centre of mass frame,
are presented for $0 < \qsq < 49\,\gevsq$ and $0.3 < y < 0.6$.
The interpretation of the results in terms of the 
structure of the virtual photon is discussed. 
The data are best described by
QCD calculations which include a 
partonic structure of the virtual photon that evolves with $Q^2$.
\end{quotation}
\vspace*{2.0cm}
{\it Submitted to Physics Letters $\mbox{\boldmath $B$}$}  \\
\cleardoublepage
\end{center}
\end{titlepage}
\noindent
\begin{flushleft}
\input h1auts
\end{flushleft}
\begin{flushleft}
\noindent
\input h1inst
\end{flushleft}
%
\newpage

\input{new.intro.tex} 

\input{det.tex} 

\input{evsel.tex} 

\input{mc.tex}

\input{results.tex}

\input{conc.tex}

\section*{Acknowledgements}

We are grateful to the HERA machine group whose outstanding
efforts have made and continue to make this experiment possible. 
We thank
the engineers and technicians for their work in constructing and now
maintaining the H1 detector, our funding agencies for 
financial support, the
DESY technical staff for continual assistance, 
and the DESY directorate for the
hospitality which they extend to the non-DESY 
members of the collaboration.


\newpage
\input{piccy.tex}
\end{document}

%% file: h1auts.tex
 C.~Adloff$^{35}$,                
 S.~Aid$^{13}$,                   
 M.~Anderson$^{23}$,              
 V.~Andreev$^{26}$,               
 B.~Andrieu$^{29}$,               
 V.~Arkadov$^{36}$,               
 C.~Arndt$^{11}$,                 
 I.~Ayyaz$^{30}$,                 
 A.~Babaev$^{25}$,                
 J.~B\"ahr$^{36}$
% ...

%% file: h1inst.tex
 $ ^1$ I. Physikalisches Institut der RWTH, Aachen, Germany$^ a$ \\
 $ ^2$ III. Physikalisches Institut der RWTH, Aachen, Germany$^ a$ \\
 $ ^3$ School of Physics and Space Research, University of Birmingham,
                             Birmingham, UK$^ b$\\
 $ ^4$ Inter-University Institute for High Energies ULB-VUB, Brussels;
   Universitaire Instelling Antwerpen, Wilrijk; Belgium$^ c$ \\
 $ ^5$ Rutherford Appleton Laboratory, Chilton, Didcot, UK$^ b$ \\
 $ ^6$ Institute for Nuclear Physics, Cracow, Poland$^ d$  \\
 $ ^7$ Physics Department and IIRPA,
         University of California, Davis, California, USA$^ e$ \\
 $ ^8$ Institut f\"ur Physik, Universit\"at Dortmund, Dortmund,
                                                  Germany$^ a$\\
 $ ^{9}$ DSM/DAPNIA, CEA/Saclay, Gif-sur-Yvette, France \\
 $ ^{10}$ Department of Physics and Astronomy, University of Glasgow,
                                      Glasgow, UK$^ b$ \\
 $ ^{11}$ DESY, Hamburg, Germany$^a$ \\
 $ ^{12}$ I. Institut f\"ur Experimentalphysik, Universit\"at Hamburg,
                                     Hamburg, Germany$^ a$  \\
 $ ^{13}$ II. Institut f\"ur Experimentalphysik, Universit\"at Hamburg,
                                     Hamburg, Germany$^ a$  \\
 $ ^{14}$ Max-Planck-Institut f\"ur Kernphysik,
                                     Heidelberg, Germany$^ a$ \\
 $ ^{15}$ Physikalisches Institut, Universit\"at Heidelberg,
                                     Heidelberg, Germany$^ a$ \\
 $ ^{16}$ Institut f\"ur Hochenergiephysik, Universit\"at Heidelberg,
                                     Heidelberg, Germany$^ a$ \\
 $ ^{17}$ Institut f\"ur Reine und Angewandte Kernphysik, Universit\"at
                                   Kiel, Kiel, Germany$^ a$\\
 $ ^{18}$ Institute of Experimental Physics, Slovak Academy of
                Sciences, Ko\v{s}ice, Slovak Republic$^{f,j}$\\
 $ ^{19}$ School of Physics and Chemistry, University of Lancaster,
                              Lancaster, UK$^ b$ \\
 $ ^{20}$ Department of Physics, University of Liverpool,
                                              Liverpool, UK$^ b$ \\
 $ ^{21}$ Queen Mary and Westfield College, London, UK$^ b$ \\
 $ ^{22}$ Physics Department, University of Lund,
                                               Lund, Sweden$^ g$ \\
 $ ^{23}$ Physics Department, University of Manchester,
                                          Manchester, UK$^ b$\\
 $ ^{24}$ CPPM, Universit\'{e} d'Aix-Marseille II,
                          IN2P3-CNRS, Marseille, France\\
 $ ^{25}$ Institute for Theoretical and Experimental Physics,
                                                 Moscow, Russia \\
 $ ^{26}$ Lebedev Physical Institute, Moscow, Russia$^ f$ \\
 $ ^{27}$ Max-Planck-Institut f\"ur Physik,
                                            M\"unchen, Germany$^ a$\\
 $ ^{28}$ LAL, Universit\'{e} de Paris-Sud, IN2P3-CNRS,
                            Orsay, France\\
 $ ^{29}$ LPNHE, Ecole Polytechnique, IN2P3-CNRS,
                             Palaiseau, France \\
 $ ^{30}$ LPNHE, Universit\'{e}s Paris VI and VII, IN2P3-CNRS,
                              Paris, France \\
 $ ^{31}$ Institute of  Physics, Czech Academy of Sciences of the
                    Czech Republic, Praha, Czech Republic$^{f,h}$ \\
 $ ^{32}$ Nuclear Center, Charles University,
                    Praha, Czech Republic$^{f,h}$ \\
 $ ^{33}$ INFN Roma~1 and Dipartimento di Fisica,
               Universit\`a Roma~3, Roma, Italy   \\
 $ ^{34}$ Paul Scherrer Institut, Villigen, Switzerland \\
 $ ^{35}$ Fachbereich Physik, Bergische Universit\"at Gesamthochschule
               Wuppertal, Wuppertal, Germany$^ a$ \\
 $ ^{36}$ DESY, Institut f\"ur Hochenergiephysik,
                              Zeuthen, Germany$^ a$\\
 $ ^{37}$ Institut f\"ur Teilchenphysik,
          ETH, Z\"urich, Switzerland$^ i$\\
 $ ^{38}$ Physik-Institut der Universit\"at Z\"urich,
                              Z\"urich, Switzerland$^ i$ \\
\smallskip
 $ ^{39}$ Institut f\"ur Physik, Humboldt-Universit\"at,
               Berlin, Germany$^ a$ \\
 $ ^{40}$ Rechenzentrum, Bergische Universit\"at Gesamthochschule
               Wuppertal, Wuppertal, Germany$^ a$ \\
 
 
\bigskip
 $ ^a$ Supported by the Bundesministerium f\"ur Bildung, Wissenschaft,
        Forschung und Technologie, FRG,
        under contract numbers 7AC17P, 7AC47P, 7DO55P, 7HH17I, 7HH27P,
        7HD17P, 7HD27P, 7KI17I, 6MP17I and 7WT87P \\
 $ ^b$ Supported by the UK Particle Physics and Astronomy Research
       Council, and formerly by the UK Science and Engineering Research
       Council \\
 $ ^c$ Supported by FNRS-NFWO, IISN-IIKW \\
 $ ^d$ Partially supported by the Polish State Committee for Scientific 
       Research, grant no. 2P03B 055 13 \\
 $ ^e$ Supported in part by USDOE grant DE~F603~91ER40674 \\
 $ ^f$ Supported by the Deutsche Forschungsgemeinschaft \\
 $ ^g$ Supported by the Swedish Natural Science Research Council \\
 $ ^h$ Supported by GA \v{C}R  grant no. 202/96/0214,
       GA AV \v{C}R  grant no. A1010619 and GA UK  grant no. 177 \\
 $ ^i$ Supported by the Swiss National Science Foundation \\
 $ ^j$ Supported by VEGA SR grant no. 2/1325/96 \\

%% file: new.intro.tex
\section{Introduction and Motivation}

In this paper jet production in electron-proton scattering in  
the transition region between photoproduction
and deep inelastic scattering (DIS) is investigated. 
The results are interpreted in terms 
of parton densities of the virtual photon 
which are probed at a scale determined by the transverse momentum
($p_t$) of the jets and which evolve with 
the virtuality of the photon ($Q^2$).

In the photoproduction of jets~\cite{expjets1}, 
the photon can couple directly to a parton
from the proton (``direct'' interactions).
However, the cross-section is dominated by interactions,
so-called ``resolved'' processes,
in which the photon fluctuates into 
a system of partons and one of these interacts with a parton out of the
proton to produce the jets. This separation into direct and 
resolved processes can only be made unambiguously in
leading order~\cite{Chyla}. 
At its simplest, the  hadronic fluctuation of the photon 
may take the form of a quark--antiquark 
($q\bar{q}$) pair. More complicated structure is built up through
QCD interactions. In addition to this point-like ``anomalous'' 
component~\cite{Witten}, the photon
can also acquire a more conventional hadronic 
structure, often modelled as a fluctuation
into a vector meson (vector dominance model, VDM). 
The cross-section for jet production can be
expressed as a convolution of
universal parton
densities of the proton and the photon 
together with hard parton-parton
scattering cross-sections. 
The evolution 
of the photon parton densities with the scale at which they are probed
can be calculated in
perturbative QCD and has been extensively measured 
in two-photon interactions~\cite{ggref1} 
and recently at HERA~\cite{efff2}.


In contrast, it is usual to consider that the only contribution to 
jet production in DIS is 
from direct interactions with the partons in the proton probed by a
structureless photon at the scale $Q^2$. 
However, in the small region of phase space where high 
$p_t$ jets are produced with $p_t^2$ much larger than $Q^2$, 
it is possible that the jet production may be most easily understood
in terms of the partonic structure 
of the virtual photon
together with that of the proton~\cite{vpth1,vpth2,epa,kkp}.
Parton densities
within the virtual photon
are expected to be 
suppressed~\cite{vpth2,vpth3,vpth4,vpth5,vpth6} 
with increasing $Q^2$ until 
direct processes dominate at $Q^2 \sim p_t^2$.
%
Measurements of the virtual photon structure in two-photon interactions
require detection of both scattered leptons at non-zero scattering angles.
Only one such measurement has previously been made~\cite{vpref}. 
The extensive $Q^2$ range together with the large centre-of-mass
energy available at HERA enables a detailed study of 
the $Q^2$ evolution of photon structure.

After a description of the data used in this analysis,
different models are introduced which are intended to describe 
the photoproduction and deep inelastic scattering regimes.
The inclusive differential 
jet cross-sections are then presented
as a function of jet transverse energy and rapidity for 
photon virtualities in the range $0 < Q^2 < 49\,\gevsq$.
Finally, a measurement of the energy flow in the direction of the
photon is shown as a function of $Q^2$ to verify the existence of 
a photon remnant. 


%% file: det.tex
\section{The H1 Detector}

A full description of the H1 detector can be 
found elsewhere~\cite{h1det} and only 
those components relevant for this analysis
are described here.

The coordinate system used has the nominal interaction point
as the origin and the incident proton beam defining the $+z$ direction.
The polar angle $\theta$ is defined with respect to the proton
direction, and the 
pseudorapidity is given by $\eta = -\ln\tan(\theta/2)$.


A finely grained Liquid Argon calorimeter~\cite{Lar}
covers the range in polar angle
$4^\circ < \theta <  154^\circ$, 
with full azimuthal
acceptance. It consists of an electromagnetic section with 
lead absorbers, 20--30 radiation lengths in depth,
and a hadronic section with steel absorbers. The
total depth of the calorimeter ranges
from 4.5 to 8 hadronic interaction lengths. 
The energy resolution is  $\sigma(E) / E \approx
0.11/ \sqrt E$ for electrons and $\sigma(E) / E \approx 0.5 / \sqrt E $
for pions ($E$ in GeV), as measured in test beams~\cite{tbeam}. 
The absolute energy scale is known to a 
precision of  3\% for electrons and 4\% for hadrons.


A series of interleaved drift and multiwire proportional chambers
surround the interaction point,
enabling the reconstruction of 
charged particles 
in the range $7^\circ < \theta <  165^\circ$, 
and the determination of the event vertex.
A uniform axial magnetic field of $1.15\,{\rm T}$ 
is provided by a superconducting coil
which surrounds the calorimeter.


For 1994 data taking, the polar region 
$151^\circ < \theta <  176^\circ$ was covered by the BEMC~\cite{BEMC}, 
a lead/scintillator
electromagnetic calorimeter with a depth of 21.7 radiation lengths.
The resolution was given by 
$0.10/\sqrt{E}$  ($E$ in GeV) and the absolute electromagnetic
energy scale was known to a precision of about 1\%.
In 1995, the BEMC was replaced by the SPACAL~\cite{spacal}, a
lead/scintillating fiber calorimeter with both an
electromagnetic and hadronic section covering the range
$153^\circ < \theta <  177.8^\circ$.
The energy resolution 
for the electromagnetic section has been determined as 
7.5\%/$\sqrt E({\rm GeV})\oplus$2.5\% and the absolute energy scale
uncertainty is 1\% at $27.5\,{\rm GeV}$ and 3\% at $7\,{\rm GeV}$~\cite{lowq2}. 
The hadronic 
energy scale uncertainty of the measurement in the SPACAL is
presently about 10\%. 
Both calorimeter sections have a time resolution
better than $1\,{\rm ns}$, 
enabling the reduction of proton beam induced
background events.  
The BEMC and the SPACAL were used both to trigger on and to measure 
the scattered lepton
in DIS processes
for $0.65<\qsq < 49\,\gevsq$.

In 1994, the backward proportional chamber (BPC) was located 
in front of the BEMC. In 1995, this was replaced by a four module
drift chamber, the BDC~\cite{BDC}, in front of the SPACAL.
The polar angle of the scattered lepton was determined using
the event vertex and 
information from these backward tracking chambers.


The luminosity system consists of two 
crystal calorimeters
with resolution $\sigma(E)/E = 0.1 / \sqrt E$ ($E$ in GeV). The 
electron tagger is located at $z = -33\,{\rm m}$ and the photon detector at
$z = -103\,{\rm m}$. The electron tagger accepts electrons with an energy
of between 0.2 and 0.8 of the incident beam energy, and
with scattering angles $\theta^{\prime} < 5\,{\rm mrad}$ 
($\theta^{\prime} = \pi - \theta$).


%% file: evsel.tex
\section{Data Samples and Event Selection}

The analysis is based on data taken by the H1 experiment 
in 1994 and 1995. Three data samples were used, each
restricted to a region of \qsq with good 
acceptance: 

\begin{itemize}
\item 
\qsq $< 10^{-2}\,\gevsq$:
The photoproduction sample in
which the scattered electron is detected in the 
electron tagger of the luminosity system. 
The events were triggered by demanding an energy
deposit in the electron tagger with $E > 4\,{\rm GeV}$
in coincidence with at least one track
pointing to the vertex region 
($p_t \ \gapprox \ 500\,{\rm MeV}$). More details of 
the trigger conditions can be found in~\cite{gptrig}.
The data sample used was a sub-sample of that 
collected in 1994 and corresponds
to an integrated luminosity of 210 ${\rm nb}^{-1}$. 
\item 
$0.65 < \qsq < 20\,\gevsq$:
The 1995 shifted vertex data sample,  
corresponding to 120 ${\rm nb}^{-1}$
collected in a special run in which the 
mean position of the 
interaction point was shifted
by $70\,{\rm cm}$ in the $+z$ direction, enabling positron
detection in the SPACAL down to angles of $178.5^\circ$. 
The events were triggered by requiring a cluster of more than $5\,{\rm GeV}$
in the SPACAL and timing consistent with an $ep$ bunch crossing.
The most energetic cluster in the electromagnetic section of the
SPACAL was taken as the electron candidate~\cite{lowq2}. 
\item 
$9< \qsq < 49\,\gevsq$: The standard 1994 data sample,
corresponding to
an integrated luminosity of 2 ${\rm pb}^{-1}$. 
The events were triggered by requiring a cluster of more than 
$4\,{\rm GeV}$ in the BEMC, and the scattered positron was 
taken as the 
most energetic BEMC cluster~\cite{elid1}. 
\end{itemize}

The error on the luminosity determination was 1.5(3)\% for data
taken in 1994(1995). 
For all three samples, the $z$ position of the interaction vertex 
was required to be within 30$\,{\rm cm}$ of the nominal position.
In addition, to enable comparisons between the DIS and photoproduction
data, for all data samples the 
inelasticity variable $y$ was restricted to 
$0.3 < y < 0.6$, the range 
where the acceptance of the electron tagger is well-understood.

For events with $0.65< \qsq < 49\,\gevsq$ it was 
also required 
that the $\sum_i(E_i-P_{zi})$ 
of all the reconstructed calorimeter 
clusters, which should be equal to twice the electron beam energy
for a DIS event,
was greater than $45\,{\rm GeV}$.
Monte Carlo studies showed that 
the background from real photoproduction events 
where hadronic activity in the 
backward region fakes a scattered lepton was reduced 
to less than 3\% in the selected kinematic region. 


The event kinematics were reconstructed from the scattered
electron 4--vector. 
Jets were reconstructed in the \gp centre of mass frame
using a $k_T$ clustering algorithm~\cite{jetalg}.  
The merging procedure
is based on the quantity $y_{ki}$ which is evaluated for
each pair of clusters:
\begin{eqnarray}
  y_{ki} &=& \frac{2(1-\cos\theta_{ki})}{E_{cut}^2}\min(E_k^2,E_i^2)
\end{eqnarray}
where $E_k$ and $E_i$ are the energies of the clusters and  
$\theta_{ki}$ is the angle between them.
In addition, to enable the association of particles   
to either a photon or proton remnant, 
two infinite momentum pseudo--particles along $\pm z$ are included
in the clustering procedure, but excluded from the final jets.
Particles are combined by the addition of their 4--vectors when 
$y_{ki}<1$. 
Thus $E_{cut}$ sets the scale for the jet resolution and 
separates the hard jets from the beam remnants. 
In this analysis, $E_{cut}$ 
was chosen to be $3\,{\rm GeV}$.
%


Jets were accepted with  
a transverse energy $\et > 4\,{\rm GeV}$ 
and a pseudorapidity 
in the range $-2.5 < \rap <-0.5$, 
where $\et$ and $\rap$ are calculated
in the $\gp$ frame with positive $\rap$ corresponding to the incident 
proton direction. 
For the photoproduction sample, the $\et$ threshold was raised to
$5\,{\rm GeV}$ to reduce the influence of
multiple parton interactions.  
This selection ensures that the energy flow around
the jet axis is well described by 
the Monte Carlo models 
used for the acceptance corrections. 


%% file: mc.tex
\section{QCD Motivated Calculations}
\label{sec:mc}

For acceptance corrections and comparisons with the 
measured jet cross-sections,
several event generators were used. 

The PHOJET 1.03~\cite{phojet} generator simulates
all relevant
components of the total photoproduction cross-section.
It is based on the two-component Dual Parton Model~\cite{DPM} and
incorporates both hard and multiple soft interactions. The 
photon flux is calculated
using the Weizs{\"a}cker--Williams approximation~\cite{WWapp,Budnev}
and the hard processes are calculated using leading order QCD matrix
elements. Final state QCD
radiation and fragmentation effects are implemented using
the string fragmentation model JETSET 7.4~\cite{jetset}. 
In this analysis, PHOJET
was used to simulate quasi-real photon-proton interactions.

DIS events were modelled using the LEPTO 6.5~\cite{lepto}
and ARIADNE 4.08~\cite{ariadne} programs.  
LEPTO includes the first order
QCD matrix elements and 
uses leading-log parton showers to model higher
order radiation.  The ARIADNE generator uses the Colour
Dipole model~\cite{cdm} to simulate QCD radiation to all orders. A
feature of this model is that the hard subprocess need 
not be generated at the photon vertex, and this  can be
regarded as generating ``resolved-like'' events. For
both models, hadronisation is performed using JETSET.




HERWIG 5.9~\cite{herwig}
was used to model direct and resolved photon processes.
The emission of the photon from the incident electron 
is generated according to the equivalent photon 
approximation~\cite{Budnev}. 
The parton-parton interactions are 
simulated according to leading order QCD calculations, 
and a parton shower 
model which effectively includes 
interference effects between the initial and
final state showers (colour coherence) is implemented ~\cite{herps}. 
The factorisation scales were 
set according to the transverse momentum of the scattered partons, 
with a cut-off at $p_t^{min}=1.5\,{\rm GeV}$. 
A cluster model is used for hadronisation.

HERWIG includes the option of additional interactions of the beam
remnant in the phenomenological soft underlying event model.
A reasonable description of the jet profiles 
and the jet rates observed 
in the data was obtained 
with no soft underlying event for $Q^2>0.65\,\gevsq$ 
and with a soft underlying event in 15\% of the 
resolved interactions at $Q^2=0\,\gevsq$. These values were
used throughout the subsequent analysis.

The RAPGAP Monte Carlo~\cite{rapgap}
originally developed to simulate diffractive
processes, also includes modeling of 
deep-inelastic and all relevant resolved photon processes.
DIS processes
are simulated using leading order QCD matrix elements 
with a $p_t^2$ cut-off scheme for the 
light quarks and the full matrix element for
heavy quarks. 
For resolved photon processes, the
equivalent photon approximation is used to
model the flux of virtual photons. 
Parton-parton interactions are calculated from
on-shell matrix elements supplemented by initial 
and final state parton
showers.

Both HERWIG and RAPGAP include models for the evolution
of the photon parton densities with $Q^2$.
Three approaches are considered.
The first assumes no $\qsq$ dependence of the parton densities.
The second uses the Drees-Godbole parameterization~\cite{vpth5}
of virtual photon structure,
following an analysis of Borzumati and Schuler~\cite{vpth6}, in
which the quark densities 
$f_{q|\gamma^*}$
in a photon of virtuality \qsq
probed at a scale $p_t^2$ 
are related to those of a real photon $f_{q|\gamma}$
by:
\begin{eqnarray}
  \label{gamsup}
  f_{q|\gamma^*}(x,p_t^2,\qsq)&=&
  f_{q|\gamma}(x,p_t^2)\;L(p_t^2,\qsq,\omega) \\
  &=& f_{q|\gamma}(x,p_t^2)
  \frac{\ln\left\{(p_t^2 + \omega^2)/(\qsq + \omega^2) \right\}}
  {\ln\left\{(p_t^2 + \omega^2)/\omega^2 \right\}}
\end{eqnarray}
An analogous relation exists for the gluon
density, with $L$ replaced by $L^2$. The parameter $\omega$ 
is the value of \qsq above which the suppression becomes significant.
We use a value of $\omega^2 = 1\,\gevsq$ and the 
GRV-G HO (DIS)~\cite{GRVgam} parameterizations for the 
unsuppressed photon parton densities.  
Throughout this paper we 
refer to this as the DG model.
The third approach is to use the photon parton densities from
Schuler and Sj$\ddot{\rm o}$strand~\cite{SAS} (SaS) 
which are valid for 
$\qsq \geq 0 \,\gevsq$. 
In this scheme, the photon parton densities are decomposed into
a direct, a VDM and a perturbative anomalous component. 
We will show comparisons with the 
SaS-2D parameterisation in the DIS scheme
using the form for the 
\qsq suppression recommended by the authors.

%

All models were used with the 
GRV94 HO (DIS)~\cite{GRVp} parton densities for
the proton, which give a good description 
of the measured $F_2$ for \qsq $> 1\,\gevsq$~\cite{lowq2}.


%% file: results.tex
\section{Inclusive Jet Cross-Section}

\subsection{Determination}

The distributions of the jet transverse energy (\ensuremath{E_t^*}) 
and 
pseudorapidity (\ensuremath{\eta^*}) 
in the \gp centre of mass frame
were corrected bin-by-bin for detector effects using generated
events passed through a 
simulation of the H1 detector based on the 
GEANT program~\cite{h1sim}.
The bin sizes were chosen to 
keep the effects of finite
resolution and bin-to-bin migration small.
For the photoproduction data, correction factors were determined
from the HERWIG DG model, which gives a good 
description of the data. 
The model dependence was estimated by comparison with the values
obtained from PHOJET, which 
also gives a good description of the jets observed in the data.
This model dependence 
is one of the largest contributions to the 
systematic error. For the DIS data, correction factors in the
range $0.65 < \qsq < 20\,\gevsq$ were determined from the HERWIG
DG model, and for $\qsq > 20\,\gevsq$ from the HERWIG direct model,
both of which give a satisfactory description of the observed
jets in these kinematic regimes. 
LEPTO was used to
estimate the model dependence of the correction factors, which is
again one of the dominant sources of systematic error. 

The other large source of systematic error arises from the 
uncertainty in the knowledge of the  
hadronic energy scale of the Liquid Argon 
calorimeter. This has two contributions; a 
possible 3\% variation in the energy scale 
between different calorimeter modules, which
is included in the point--to--point error, and a 
4\% uncertainty in the overall energy scale, which affects the 
normalisation of the jet cross-sections.

Further sources of systematic error include a 1(2)\% uncertainty 
in the electromagnetic energy scale of the BEMC(SPACAL) and 
a 1\,mrad uncertainty in the polar angle of the scattered electron. 
For the photoproduction data, 
the uncertainty in the acceptance and energy calibration of the
electron tagger was included.
A 20(10)\% uncertainty in the knowledge of the hadronic energy 
scale of the BEMC(SPACAL) is also considered. 
The 1.5(3)\% uncertainty in the luminosity determination in 
1994(1995) affects the overall normalisation of the 
jet cross-sections.

The effect of radiative corrections 
in DIS events has been studied using
the HERACLES program~\cite{heracles} which 
includes complete first order radiative corrections and the 
emission of real bremsstrahlung photons for the 
electroweak interaction. 
The effect is 20--30\% for jets with $\et$ of $4\,{\rm GeV}$,
decreases with increasing $\et$ and is
negligible for $\et>7\,{\rm GeV}$. 
It does not significantly influence the 
conclusions and the data are not corrected for
this effect.

The corrected cross-sections obtained from the 1995
shifted vertex data and from the 1994 data 
are in good agreement in the region 
$9<\qsq<20\,\gevsq$ where the data samples overlap.

\subsection{Results}

Figure~\ref{fig:siget} shows the inclusive $ep$ jet
cross--section \dsiget for 0.3 $ < y < $ 0.6 and jets
with $-2.5 < \eta^* < -0.5$, and the values are
listed in table~\ref{tab:xet}.
The data are compared with 
the prediction from the HERWIG DG model, which 
includes a resolved component of the virtual photon.
This is able to give a good description of the data except
for jets in the lowest $\et$ range 
when  $9 < \qsq < 49\,\gevsq$.
Also shown is the direct contribution to this model
which accounts for an increasing fraction of the total
prediction as \qsq increases, but which alone
cannot describe the measured jet cross-sections. 



%
%

The jet cross-section \dsigrap for jets 
with \et $> 5\,{\rm GeV}$ is shown in
figure~\ref{fig:sigeta} 
and the values are listed in table~\ref{tab:xeta}.
The data are compared to the HERWIG DG model and the direct 
photon contribution to this model. 
The direct photon processes alone
significantly underestimate the 
jet cross-section at low $Q^2$, but the data
are described by HERWIG if the resolved photon 
component is included and suppressed with increasing $Q^2$
according to the DG model.
The relative contribution to the jet cross-section
from resolved photon processes
increases towards the proton (+\ensuremath{\eta^*}) direction.  
%
%

In order to study in more detail the \qsq evolution of the photon
parton densities, we factor out the \qsq dependence contained in
the flux of photons 
and calculate a \gp jet cross-section in each \qsq range using:
\begin{gather}
\sigma_{\gamma^*p\rightarrow jet+X}
=\frac{\sigma_{ep \rightarrow e+jet+X}}
{F_{\gamma|e}}\label{eq:f6}
\end{gather}
We use the Weizs$\ddot{\rm a}$cker--Williams 
approximation~\cite{WWapp, Budnev}
to calculate the flux of photons, $F_{\gamma|e}$:
\begin{gather}
F_{\gamma|e}=
\int_{y_{min}}^{y_{max}}{\rm d}y\int_{Q^2_{min}}^{Q^2_{max}}
{\rm d}Q^2 f_{\gamma|e}(y,Q^2)\label{eq:f3}\\ 
\intertext{with}
f_{\gamma |e}(y,Q^2)=\frac{\alpha}{2\pi Q^2}\left\{
  \frac{1+(1-y)^2}{y}-\frac{2(1-y)}{y}\frac{Q^2_{min}}{Q^2}\right\} \label{eq:f4
}\\
\intertext{The flux is integrated over $0.3 < y < 0.6$ and
$Q^2_{max}$ and $Q^2_{min}$ are the upper and lower edges of the \qsq range.
For photoproduction:}
Q_{min}^2=\frac{m_e^2y^2}{1-y} \label{eq:f5}
\end{gather}
where $m_e$ is the electron mass.

This factorisation of the cross-section remains a 
reasonable approximation 
for high $p_t$ jet production 
if $p_t^2 \gg \qsq$~\cite{vpth2}, a condition which
is satisfied by the majority of our data\footnote{ 
We continue to use (\ref{eq:f6}) as a formal definition of the
\gp cross-section for all $p_t^2$ and $Q^2$.}.
The numerical values of the flux factors used in each $Q^2$ range 
are listed in table~\ref{tab:flux}.

\begin{table}[tb]
\begin{center} 
\begin{tabular}{|c|c|}
  \hline
  \qsq (${\rm GeV}^2$) & Flux factor \ \\
  \hline\hline
  $< 10^{-2}$ & $1.16\times 10^{-2}$ \ \\
  \hline
  $0.65< \qsq < 1.2$ & $6.56\times 10^{-4}$ \ \\
  \hline
  $1.2 < \qsq <2.6$ & $8.27\times 10^{-4}$ \ \\
  \hline
  $2.6 < \qsq < 4$ & $4.61\times 10^{-4}$ \ \\
  \hline
  $4 < \qsq < 9$ & $8.68\times 10^{-4}$ \ \\
  \hline
  $9 < \qsq < 20$ & $8.54\times 10^{-4}$ \ \\
  \hline
  $20 < \qsq < 25$ & $2.39\times 10^{-4}$ \ \\
  \hline
  $25 < \qsq < 36$ & $3.90\times 10^{-4}$ \ \\
  \hline
  $36 < \qsq < 49$ & $3.30\times 10^{-4}$ \ \\
  \hline
\end{tabular}
\end{center} 
\caption[]{The flux factors used for the calculation of the
\gp jet cross-sections (see equation~\ref{eq:f3}).} 
\label{tab:flux}   
\end{table}

The \qsq dependence of the 
inclusive \gp jet cross-section at fixed jet \et is shown in
figure~\ref{fig:dis}.
For \et $< 10\,{\rm GeV}$ there is a significant decrease of the 
jet cross-section with increasing $Q^2$.
Also shown for comparison are the predictions from LEPTO
and ARIADNE. We expect such DIS models to be valid when
$\qsq \ \gapprox \ \ensuremath{E_t^{*2}}$, 
where the photon cannot be resolved.
This region corresponds to the 2--3 highest \qsq ranges in 
figures~\ref{fig:dis}a and~\ref{fig:dis}b.  
It can be seen that both models 
give an adequate description of the data in these ranges.
However, neither model can describe the data when 
$\qsq < \ensuremath{E_t^{*2}}$ and the virtual photon can be
resolved. 
Also the predictions of these models 
differ significantly in this
region. 
Although ARIADNE predicts
a jet cross-section which decreases with increasing \qsq
in a similar manner to the data
and is able to describe the data for $\qsq \ \gapprox \  4\,{\rm GeV}^2$, 
it is unable to describe the 
data at all \et and all $Q^2$.
The prediction from ARIADNE is sensitive to the 
parameters which limit
the phase space for QCD radiation\footnote{PARA(10) and PARA(15).
 We use PARA(10)=1.5 and PARA(15)=0.5 by default.}
and therefore to the fraction of
``resolved-like'' events produced by the model, 
but we found no choice of
these parameters which enabled the model to describe the 
data. 

%
%

The same data compared to 
a series of models which 
include a partonic structure for the virtual photon, 
as implemented in the HERWIG and RAPGAP generators, are shown in
 figure~\ref{fig:gamstr}.
In each case, the sum of the direct and resolved contributions is
shown. 
The dot--dashed curve shows the prediction 
from HERWIG assuming the GRV-G HO 
structure function for the photon with no \qsq suppression. 
The resulting \gp 
jet cross-section is 
almost independent of $Q^2$, in contrast to the
data which show, except for the highest \et jets, a significant
decrease of the jet cross-section with $Q^2$.
The jet cross-section predicted by this model
at $\qsq=0\,\gevsq$ is slightly larger than that predicted for
$\qsq>0\,\gevsq$ because a soft underlying event is included
at $\qsq=0\,\gevsq$.   
Also shown are the predictions from HERWIG and from RAPGAP using 
the DG model for the virtual photon structure. 
The models are in good agreement with each other, and describe the
data well except for jets with
$4<\et<5\,{\rm GeV}$ 
when  $9 < \qsq < 49\,\gevsq$,
where they 
underestimate the measured cross-section.
We note that in the DG model
the photon parton density functions 
all vanish for $\qsq > p_t^2$,
which is approximately the case in  
this region.
The data are best described by the RAPGAP model using the 
SaS-2D parameterization of the virtual photon. In contrast
to the DG model, the 
photon parton densities do not vanish when $\qsq > p_t^2$ in this
parameterization.  

The inclusive jet cross-section can
therefore be understood if a partonic 
structure is ascribed to the virtual photon.
Moreover, the observed \qsq evolution of the jet cross-sections
can be explained by the suppression of the parton densities
with increasing photon virtuality as
predicted by QCD inspired models. For $\qsq \ \gapprox \ p_t^2$, the
photon is effectively structureless.

\section{Photon Remnant}

The jet algorithm used in this analysis assigns particles
to a photon remnant.
The fraction of the incident photon's energy which is
reconstructed in the photon remnant jet is given by:
\begin{eqnarray*}
  f = \frac{\sum_i E_i^*}{E_e^* - E_e^{\prime*}}  
\end{eqnarray*}
where $E_i^*$ is the energy of a particle assigned to the photon 
remnant
and $E_e^*$ and $E_e^\prime*$ are the energies of the  
incident and scattered electron respectively 
in the \gp frame.
 
Figure~\ref{fig:reme} shows the uncorrected distribution of
$f$ in the data, as a function of $Q^2$, for events with at least
one jet with \et $> 5\,{\rm GeV}$ and $-2.5 < \eta^* < -0.5$. 
At $\qsq = 0\,\gevsq$, where resolved photon processes dominate
the cross-section, most of the events with jets also contain a 
photon remnant with a significant fraction of the incident photon's
energy. Conversely, at the highest \qsq values, where the direct
processes dominate, $f$ is peaked at zero. 
Also shown are the predictions from the HERWIG
DG model and from LEPTO after detector simulation. It can be
seen that the distribution of $f$ from LEPTO is peaked at zero
for all $Q^2$, as expected for a model which includes only direct
processes. 
At low $Q^2$, the data agree with the HERWIG DG model, and at
high \qsq they agree with the LEPTO prediction.

The evolution of $f$ with \qsq is consistent with the picture of
a resolved photon contribution which is suppressed 
with increasing virtuality.
%


%% file: conc.tex
\section{Conclusions}

The inclusive $ep$ jet cross-sections \dsiget and \dsigrap
have been measured
in the kinematic range 0.3 $<y<$ 0.6 and 0 $< Q^2 <$ 49 \gevsq.

Models in which the photon only 
couples directly to the partons of the hard scattering process
fail to describe the data in the region
$\ensuremath{E_t^{*2}}\ \gapprox \ \qsq$, 
where the virtual photon can be resolved.
Models which include a resolved
component of the photon suppressed with \qsq 
are in good agreement with the data.
The best description of the data was obtained with a model
which includes direct, VDM and perturbative contributions to
the virtual photon structure.

It has been established that the energy assigned to
the photon remnant 
in events with high $\et$ jets
is on average large at low \qsq and decreases 
with increasing $Q^2$, consistent with
the picture of a resolved photon component which is suppressed 
with its increasing virtuality.


%% file: piccy.tex
\begin{figure}[h]
\begin{center}
\epsfig{file=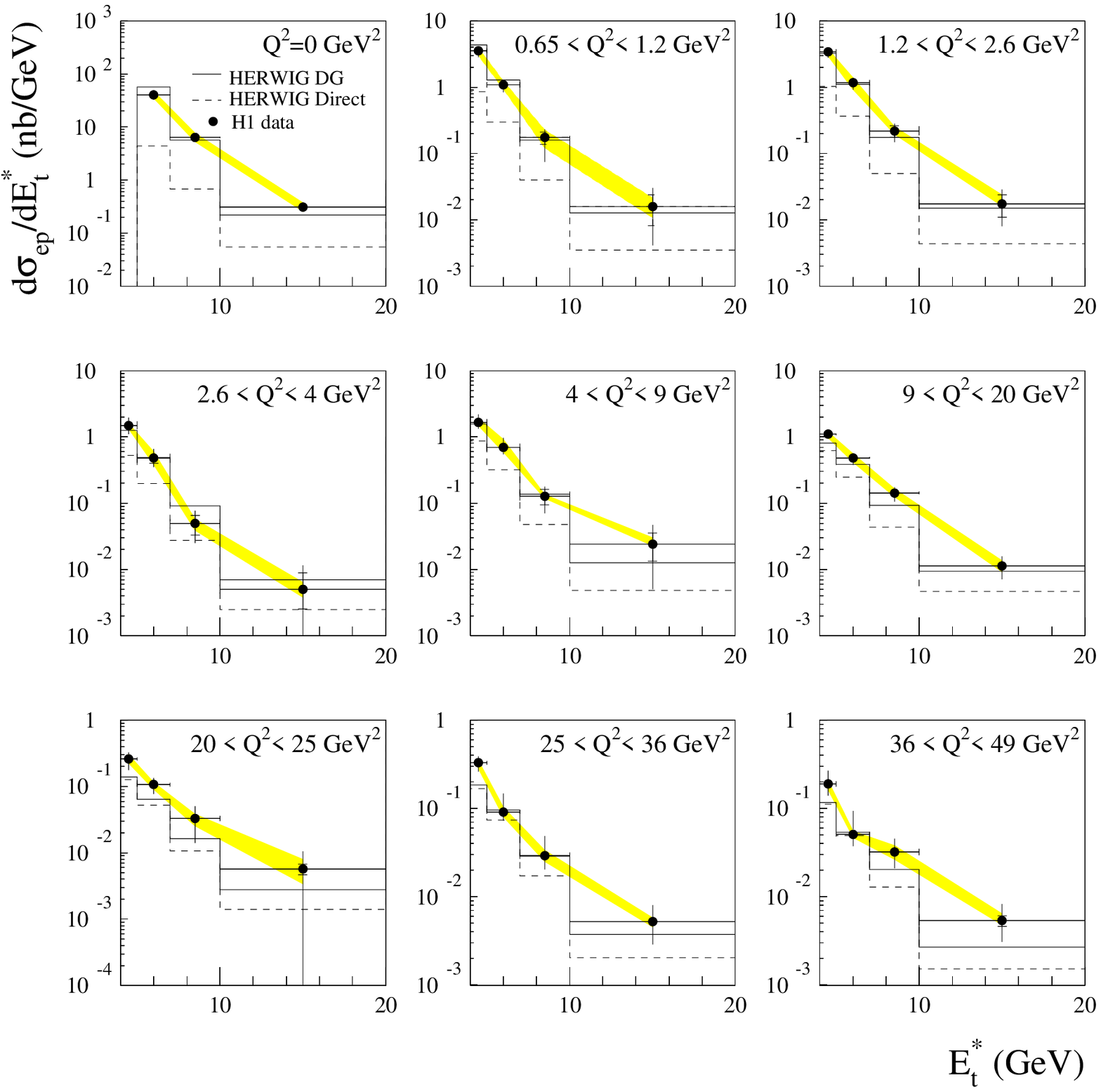,width=\textwidth}
\caption{The differential jet cross-section 
\dsiget for jets with $-2.5< \rap < -0.5$ and  $0.3 < y < 0.6$.
The inner error bars indicate the
statistical errors, the total error bars show the quadratic sum
of the statistical and systematic errors 
and the shaded band represents the correlated
error from the uncertainty in the Liquid Argon energy scale. 
Not shown is the 
error from the uncertainty in the luminosity determination 
which leads to a
3\% normalisation error for the
data with $0.65<\qsq<9\,{\rm GeV}^2$ 
and a 1.5\% normalisation error elsewhere.
The data are compared to
the HERWIG DG model (solid line) 
and to the direct contribution to this model (dashed line).}
\label{fig:siget}
\end{center} 
\end{figure}
\newpage

\begin{figure}[h]
\begin{center}
\epsfig{file=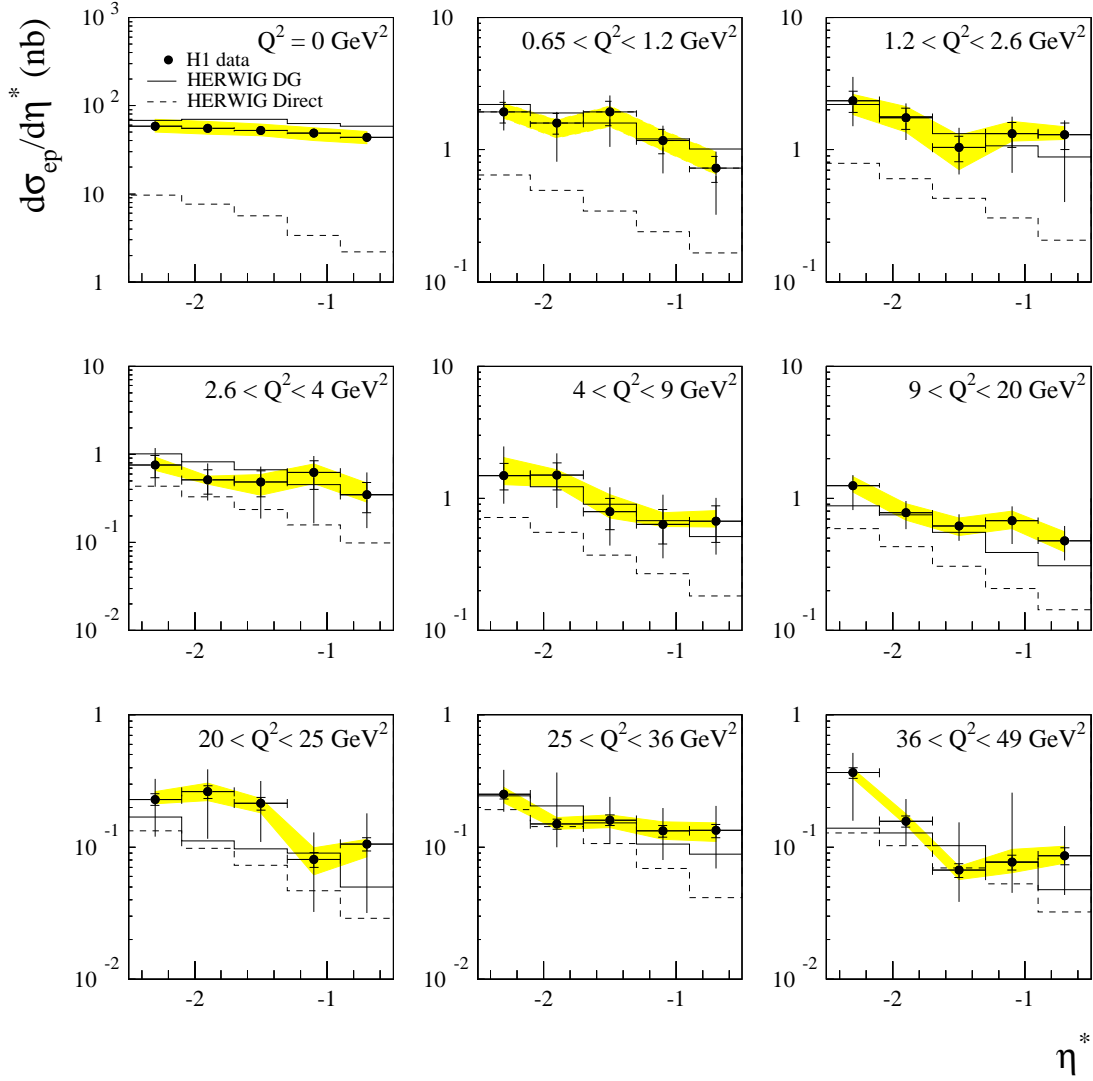,width=\textwidth}
\caption{The differential jet cross-section 
\dsigrap for jets with $\et > 5\,\gevsq$ and  $0.3 < y < 0.6$. The
incident proton direction is to the right. 
The inner error bars indicate the
statistical errors, 
the total error bars show the quadratic sum
of the statistical and systematic errors 
and the shaded band represents the correlated
error from the uncertainty in the Liquid Argon energy scale.
Not shown is the 
error from the uncertainty in the luminosity determination 
which leads to a
3\% normalisation error for the
data with $0.65<\qsq<9\,{\rm GeV}^2$
and a 1.5\% normalisation error elsewhere.
The data are compared
to the HERWIG DG model (solid line) 
and to the direct contribution to this model (dashed line). }
\label{fig:sigeta}
\end{center} 
\end{figure}
\newpage

\begin{figure}[h]
\begin{center}
\epsfig{file=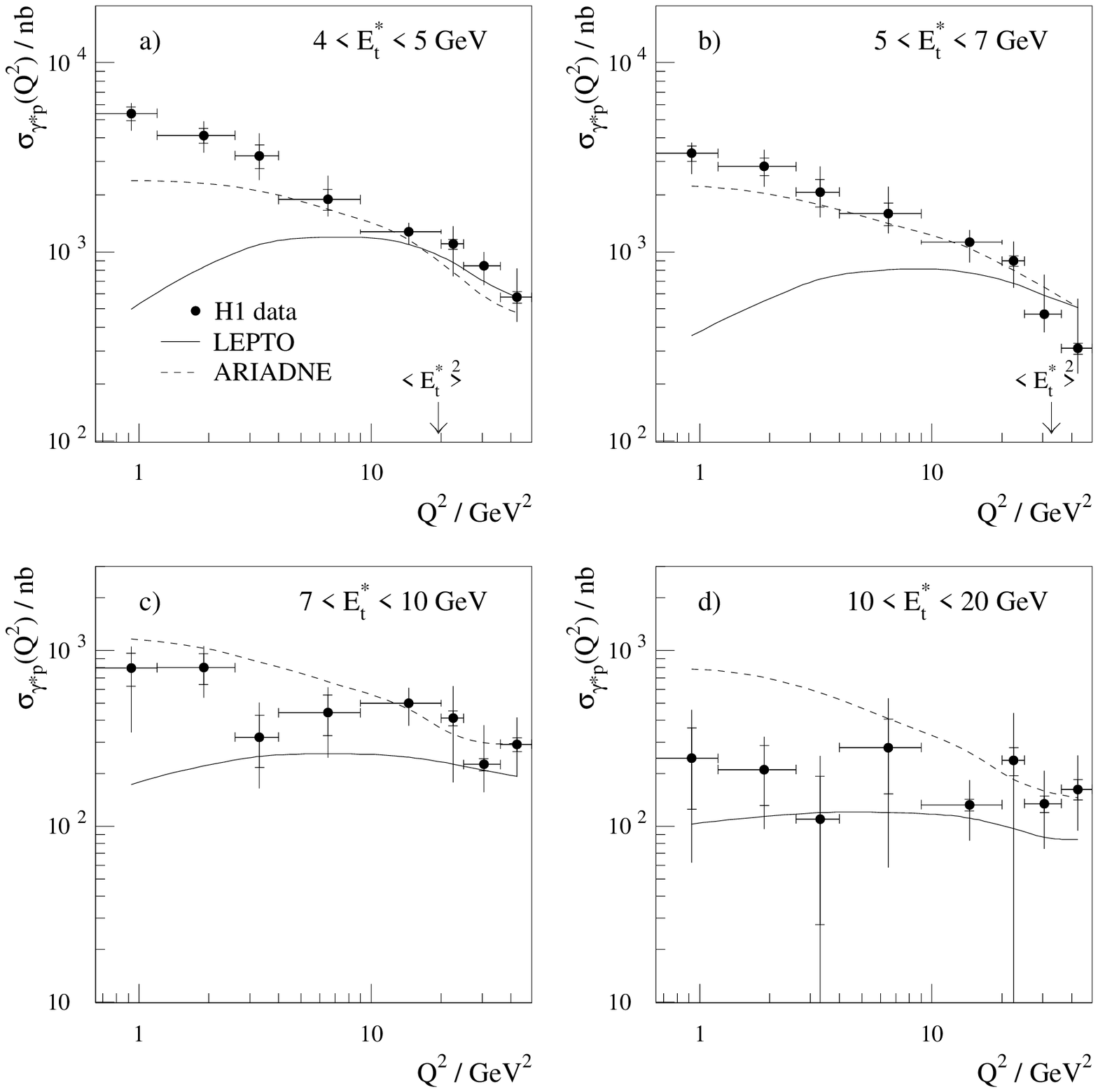,width=\textwidth}
\caption{The inclusive \gp jet cross-section 
$\sigma_{\gp}(Q^2)$ 
for jets with $-2.5< \rap < -0.5$ and  $0.3 < y < 0.6$. 
The inner error bars indicate the
statistical errors and 
the total error bars show the quadratic sum
of the statistical and systematic errors.
Not shown are the normalisation error from the 
uncertainty in the Liquid Argon energy scale,
which is 15\% at low \et and increases to 25\% at
high \ensuremath{E_t^*}, and the normalisation
error from the uncertainty in the luminosity determination, 
which is 
3\% for the
data with $0.65<\qsq<9\,{\rm GeV}^2$ and 1.5\% elsewhere.
The data are compared to
LEPTO (solid line) and 
ARIADNE (dashed line).} 
\label{fig:dis}
\end{center} 
\end{figure}
\newpage

\begin{figure}[h]
\begin{center}
\epsfig{file=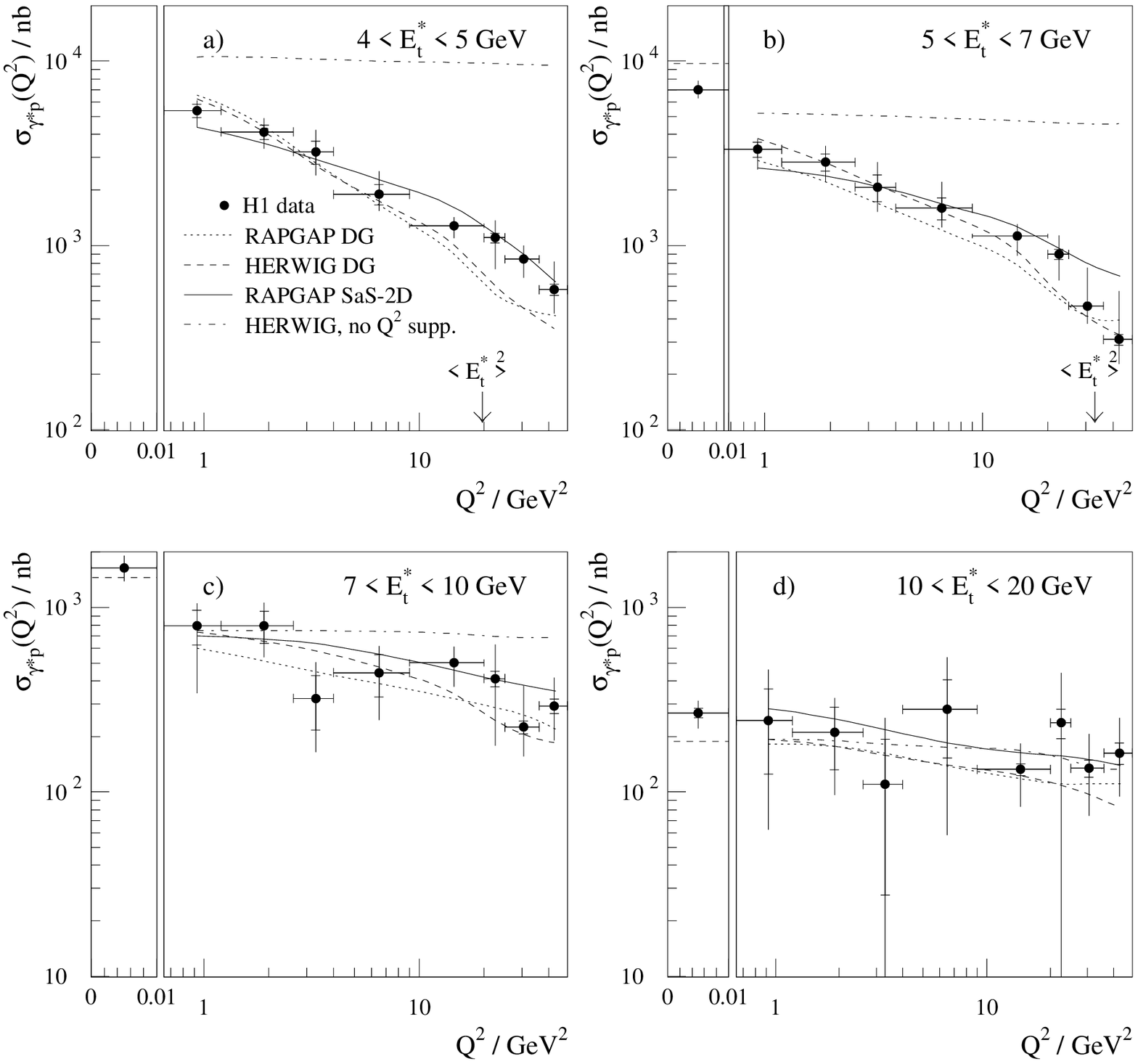,width=\textwidth}
\caption{The inclusive \gp jet cross-section $\sigma_{\gp}(Q^2)$   
for jets with $-2.5< \rap < -0.5$ and  $0.3 < y < 0.6$.
The inner error bars indicate the
statistical errors and 
the total error bar shows the quadratic sum
of the statistical and systematic errors.
Not shown are the normalisation error from the 
uncertainty in the Liquid Argon energy scale,
which is 15\% at low \et and increases to 25\% at
high \ensuremath{E_t^*}, and the normalisation
error from the uncertainty in the luminosity determination, 
which is 
3\% for the
data with $0.65<\qsq<9\,{\rm GeV}^2$ and 1.5\% elsewhere.
The data 
are compared to HERWIG with no suppression of the
photon structure function with \qsq (dot--dashed line),
the HERWIG DG model (dashed line), 
the RAPGAP DG model (dotted line)
and RAPGAP with the
SaS-2D photon structure function (solid line).}
\label{fig:gamstr}
\end{center} 
\end{figure}
\newpage

\begin{figure}[h]
\begin{center}
\epsfig{file=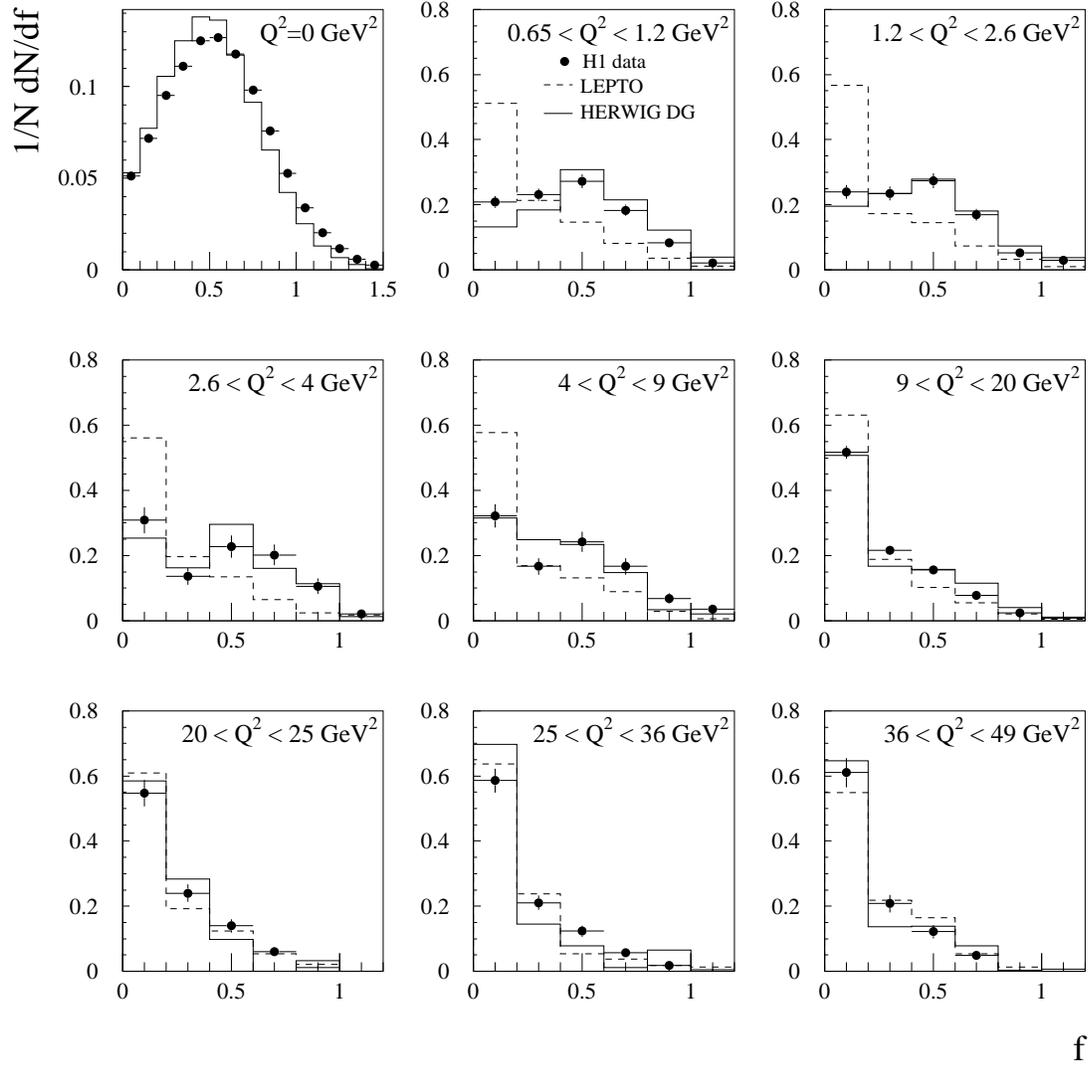,width=\textwidth}
\caption{The uncorrected
distribution of the observed fraction of the incident
photon's energy which is reconstructed in the photon remnant jet
 ($f$)  
for events with at least one jet with \et $>5\,{\rm GeV}$ and 
$-2.5 < \eta^* <  -0.5$. The data are compared to the 
HERWIG DG model (solid line) and to 
LEPTO (dashed line) after detector simulation.}
\label{fig:reme}
\end{center} 
\end{figure}
\newpage
\begin{table}[h]
\begin{center} 
\begin{tabular}{|c|c||c|c|c|c|c|}
  \hline
   \qsq  & \et 
& \dsiget & $\delta(stat)$ & $+\delta(sys)$ & $-\delta(sys)$ 
& $\pm\delta(norm)$\\
 (${\rm GeV}^2$) & (GeV) & (nb/GeV) & & & & \\
   \hline\hline
   $\qsq<10^{-2}$ & $5-7$ & 40.7 & 0.5 & 4.8 & 4.0 & 7.1 \ \\
   & $7-10$ & 6.4 & 0.2 & 1.0 & 1.0 & 1.3 \ \\
   & $10-20$ & 0.31 & 0.02 & 0.05 & 0.06 & 0.06 \ \\
  \hline
   $0.65<\qsq<1.2$ & $4-5$ & 3.5 & 0.3 & 0.4 & 0.6 & 0.6 \ \\
   & $5-7$ & 1.1 & 0.1 & 0.1 & 0.2 & 0.2 \ \\
   & $7-10$ & 0.17 & 0.04 & 0.04 & 0.09 & 0.05 \ \\
   & $10-20$ & 0.016 & 0.008 & 0.012 & 0.009 & 0.005 \ \\
  \hline
   $ 1.2<\qsq<2.6$ & $4-5$ & 3.4 & 0.3 & 0.5 & 0.6 & 0.5 \ \\
   & $5-7$ & 1.2 & 0.1 & 0.2 & 0.2 & 0.2 \ \\
   & $7-10$ & 0.22 & 0.04 & 0.04 & 0.06 & 0.03 \ \\
   & $10-20$ & 0.017 & 0.007 & 0.009 & 0.007 & 0.007 \ \\
  \hline
  $2.6<\qsq<4$ & $4-5$ & 1.5 & 0.2 & 0.4 & 0.3 & 0.2 \ \\
   & $5-7$ & 0.48 & 0.08 & 0.16 & 0.10 & 0.13 \ \\
   & $7-10$ & 0.049 & 0.016 & 0.023 & 0.018 & 0.008 \ \\
   & $10-20$ & $5.1\times 10^{-3}$ 
   & $3.8\times 10^{-3}$ & $5.3\times 10^{-3}$
   & $6.9\times 10^{-3}$ & $1.3\times 10^{-3}$\ \\
  \hline
   $4<\qsq<9$ & $4-5$ & 1.6 & 0.2 & 0.5 & 0.2 & 0.3 \ \\
   & $5-7$ & 0.69 & 0.09 & 0.26 & 0.12 & 0.22 \ \\
   & $7-10$ & 0.13 & 0.03 & 0.04 & 0.05 & 0.01 \ \\
   & $10-20$ & 0.024 & 0.011 & 0.019 & 0.016 & 0.006 \ \\
  \hline
  $9<\qsq<20$ & $4-5$ & 1.10 & 0.03 & 0.12 & 0.16 & 0.12 \ \\
   & $5-7$ & 0.48 & 0.01 & 0.07 & 0.11 & 0.08\ \\
   & $7-10$ & 0.143 & 0.007 & 0.032 & 0.037 & 0.022 \ \\
   & $10-20$ & $1.13\times 10^{-2}$ 
   & $0.09\times 10^{-2}$ 
   & $0.44\times 10^{-2}$ & $0.42\times 10^{-2}$ 
   & $0.23\times 10^{-2}$\ \\
  \hline
    $20<\qsq<25$ & $4-5$ & 0.26 & 0.02 & 0.06 & 0.08 & 0.04 \ \\
   & $5-7$ & 0.107 & 0.007 & 0.028 & 0.029 & 0.016\ \\
   & $7-10$ & 0.033 & 0.003 & 0.017 & 0.019 & 0.007\ \\
   & $10-20$ & $5.7\times 10^{-3}$ 
   & $1.0\times 10^{-3}$ & $4.9\times 10^{-3}$
   & $60.8\times 10^{-3}$ & $2.3\times 10^{-3}$\ \\
  \hline
    $25<\qsq<36$ & $4-5$ & 0.33 & 0.02 & 0.06 & 0.07 & 0.04\ \\
   & $5-7$ & 0.092 & 0.005 & 0.056 & 0.019 & 0.011\ \\
   & $7-10$ & 0.029 & 0.002 & 0.020 & 0.009 & 0.004 \ \\
   & $10-20$ & $5.2\times 10^{-3}$ & $0.6\times 10^{-3}$ 
   & $2.8\times 10^{-3}$ & $2.3\times 10^{-3}$ 
   & $0.7\times 10^{-3}$\ \\
  \hline
    $36<\qsq<49$ & $4-5$ & 0.19 & 0.01 & 0.08 & 0.05 & 0.03\ \\
   & $5-7$ & 0.051 & 0.003 & 0.042 & 0.013 & 0.004\ \\
   & $7-10$ & 0.032 & 0.003 & 0.014 & 0.011 & 0.006 \ \\
   & $10-20$ & $5.4\times 10^{-3}$ & $0.7\times 10^{-3}$ 
   & $3.0\times 10^{-3}$ & $2.3\times 10^{-3}$ 
   & $1.1\times 10^{-3}$\ \\
  \hline
\end{tabular}
\end{center} 
\caption[]{The inclusive differential jet cross-section \dsiget
for jets with $-2.5 < \rap < -0.5$ in the \gp centre of mass frame
measured in the range $0.3 < y < 0.6$ for nine different \qsq ranges.
The statistical, positive systematic, negative systematic and normalisation
errors are given. 
In addition, 
the uncertainty in the luminosity determination leads to 
a 3\% normalisation error for the data with 
$0.65<\qsq<9\,\gevsq$ and a 1.5\% normalisation error elsewhere.} 
\label{tab:xet}   
\end{table} 
\newpage

\begin{table}[h]
\begin{center} 
\begin{tabular}{|c|c||c|c|c|c|c|}
  \hline
   \qsq (${\rm GeV}^2$)  & \rap & \dsigrap (nb) & 
   $\delta(stat)$ & +$\delta(sys)$ & -$\delta(sys)$ 
& $\pm\delta(norm)$\\
   \hline\hline
    $\qsq<10^{-2}$ & $-2.5<\rap <-2.1$ & 58.6 & 1.4 & 7.9 & 6.8 & 11.6\ \\
             & $-2.1<\rap <-1.7$ & 55.8 & 1.3 & 6.8 & 5.8 & 10.3 \ \\
             & $-1.7<\rap <-1.3$ & 52.4 & 1.2 & 6.3 & 4.6 & 9.7 \ \\
             & $-1.3<\rap <-0.9$ & 48.9 & 1.1 & 5.4 & 5.5 & 8.3 \ \\
             & $-0.9<\rap <-0.5$ & 43.6 & 1.0 & 5.5 & 4.4 & 8.5 \ \\
  \hline
    $0.65<\qsq<1.2$ & $-2.5<\rap <-2.1$ & 1.9 & 0.3 & 0.8 & 0.4 & 0.3\ \\
             & $-2.1<\rap <-1.7$ & 1.6 & 0.3 & 0.2 & 0.7 & 0.4 \ \\
             & $-1.7<\rap <-1.3$ & 1.9 & 0.4 & 0.5 & 0.8 & 0.4 \ \\
             & $-1.3<\rap <-0.9$ & 1.2 & 0.2 & 0.2 & 0.4 & 0.2 \ \\
             & $-0.9<\rap <-0.5$ & 0.73 & 0.16 & 0.18 & 0.37 & 0.24 \ \\
  \hline
    $1.2<\qsq<2.6$ & $-2.5<\rap <-2.1$ & 2.3 & 0.4 & 1.1 & 0.7 & 0.5 \ \\
             & $-2.1<\rap <-1.7$ & 1.7 & 0.3 & 0.4 & 0.5 & 0.4 \ \\
             & $-1.7<\rap <-1.3$ & 1.0 & 0.2 & 0.4 & 0.3 & 0.3 \ \\
             & $-1.3<\rap <-0.9$ & 1.3 & 0.3 & 0.4 & 0.6 & 0.3 \ \\
             & $-0.9<\rap <-0.5$ & 1.3 & 0.3 & 0.2 & 0.8 & 0.2 \ \\
  \hline
    $2.6<\qsq<4$ & $-2.5<\rap <-2.1$ & 0.75 
             & 0.21 & 0.37 & 0.25 & 0.18 \ \\
             & $-2.1<\rap <-1.7$ & 0.51 & 0.16 
             & 0.23 & 0.14 & 0.05 \ \\
             & $-1.7<\rap <-1.3$ & 0.49 
             & 0.16 & 0.17 & 0.26 & 0.15 \ \\
             & $-1.3<\rap <-0.9$ & 0.62 & 0.22 
             & 0.25 & 0.40 & 0.16 \ \\
             & $-0.9<\rap <-0.5$ & 0.35 
             & 0.13 & 0.24 & 0.16 & 0.12 \ \\
  \hline
    $4<\qsq<9$ & $-2.5<\rap <-2.1$ & 1.5 & 0.3 
             & 0.9 & 0.5 & 0.5 \ \\
             & $-2.1<\rap <-1.7$ & 1.5 & 0.4 
             & 0.6 & 0.6 & 0.3 \ \\
             & $-1.7<\rap <-1.3$ & 0.79 & 0.21 
             & 0.37 & 0.28 & 0.27 \ \\
             & $-1.3<\rap <-0.9$ & 0.64 & 0.19 
             & 0.39 & 0.22 & 0.14 \ \\
             & $-0.9<\rap <-0.5$ & 0.67 & 0.21 
             & 0.27 & 0.21 & 0.13 \ \\
  \hline
    $9<\qsq<20$ & $-2.5<\rap <-2.1$ & 1.25 & 0.06
             & 0.25 & 0.44 & 0.20 \ \\
             & $-2.1<\rap <-1.7$ & 0.78 & 0.04
             & 0.18 & 0.19 & 0.14 \ \\
             & $-1.7<\rap <-1.3$ & 0.62 & 0.03
             & 0.14 & 0.14 & 0.10 \ \\
             & $-1.3<\rap <-0.9$ & 0.68 & 0.04
             & 0.19 & 0.23 & 0.12 \ \\
             & $-0.9<\rap <-0.5$ & 0.48 & 0.03 
             & 0.14 & 0.14 & 0.09 \ \\
  \hline
    $20<\qsq<25$ & $-2.5<\rap <-2.1$ & 0.23 & 0.02
             & 0.10 & 0.11 & 0.03 \ \\
             & $-2.1<\rap <-1.7$ & 0.26 & 0.03
             & 0.13 & 0.15 & 0.04 \ \\
             & $-1.7<\rap <-1.3$ & 0.21 & 0.02
             & 0.10 & 0.11 & 0.03 \ \\
             & $-1.3<\rap <-0.9$ & 0.08 & 0.01
             & 0.05 & 0.05 & 0.02 \ \\
             & $-0.9<\rap <-0.5$ & 0.11 & 0.01
             & 0.07 & 0.07 & 0.02 \ \\
  \hline
    $25<\qsq<36$ & $-2.5<\rap <-2.1$ & 0.25 & 0.02
             & 0.14 & 0.07 & 0.03 \ \\
             & $-2.1<\rap <-1.7$ & 0.15 & 0.01
             & 0.22 & 0.05 & 0.02\ \\
             & $-1.7<\rap <-1.3$ & 0.16 & 0.02
             & 0.08 & 0.05 & 0.02 \ \\
             & $-1.3<\rap <-0.9$ & 0.13 & 0.01
             & 0.07 & 0.05 & 0.02 \ \\
             & $-0.9<\rap <-0.5$ & 0.13 & 0.02
             & 0.07 & 0.06 & 0.02 \ \\
  \hline
    $36<\qsq<49$ & $-2.5<\rap <-2.1$ & 0.37 & 0.03
             & 0.15 & 0.21 & 0.03 \ \\
             & $-2.1<\rap <-1.7$ & 0.16 & 0.02
             & 0.07 & 0.06 & 0.02 \ \\
             & $-1.7<\rap <-1.3$ & 0.067 & 0.008
             & 0.087 & 0.028 & 0.010 \ \\
             & $-1.3<\rap <-0.9$ & 0.077 & 0.010
             & 0.18 & 0.032 & 0.019 \ \\
             & $-0.9<\rap <0.5$ & 0.09 & 0.01 & 0.06 & 0.04 & 0.02\ \\
  \hline
\end{tabular}
\end{center}
\caption[]{The inclusive differential jet cross-section \dsigrap
for jets with $\et > 5\,{\rm GeV}$ in the \gp centre of mass frame
measured in the range $0.3 < y < 0.6$ for nine different \qsq ranges.
The statistical, positive systematic, negative systematic
and normalisation errors are given. 
In addition, 
the uncertainty in the luminosity determination leads to 
a 3\% normalisation error for the data with 
$0.65<\qsq<9\,\gevsq$ and a 1.5\% normalisation error elsewhere.} 
\label{tab:xeta}   
\end{table}
